\documentclass[aps,prl,twocolumn,superscriptaddress,letterpaper,longbibliography]{revtex4-2}
\usepackage{graphicx}
\usepackage{bm}
\usepackage{amssymb}
\usepackage{color}
\usepackage{amsmath}
\usepackage{ulem}
\usepackage{bbold}
\usepackage[mathlines]{lineno}
\usepackage[colorlinks=True,citecolor=blue,linkcolor=blue,allcolors=blue]{hyperref}
\usepackage{enumitem}
\usepackage[utf8]{inputenc}

\begin{document}

\title{Floquet $\pi$ Exceptional Points}

\author{Weiwei Zhu}
\email{phyzhuw@ouc.edu.cn}
\affiliation{College of Physics and Optoelectronic Engineering, Ocean University of China, Qingdao 266100, China}
\affiliation{Key Laboratory for Optics Photoelectronics, Qingdao 266100, China}
\affiliation{Engineering Research Center of Advanced Marine Physical Instruments and Equipment of MOE, Qingdao 266100, China}

\begin{abstract}
We report a new kind of exceptional points in periodically driven system, called Floquet $\pi$ exceptional points, whose eigenvectors rotate on Bloch sphere and accumulate $\pi$ geometric phase in one time period. The merging of two such kind exceptional points are constrained by their dynamical structure, meaning two order-1/2 exceptional points with same dynamical structure can merge to one order-1 one while those with opposite dynamical structure can not. We show they exist in Floquet bipartite lattices, and the order-1 Floquet $\pi$ exceptional points appear at the phase transition point between quasimomentum gap phases and quasienergy gap phases. The scattering properties around the order-1 Floquet $\pi$ exceptional points is quite novel, which is perfect transparency but detectable in reflection for one of two sides.
\end{abstract}

\maketitle

{\it Introduction.--} The development of quantum mechanics is mainly based on the assumption of Hermiticity, and the non-Hermiticity is only treated as a perturbation which is usually considered to have little impact on the main properties of the system~\cite{Pauli1958,Messiah1995,Kato1966}. Recent studies on non-Hermitian physics have shown its realization on novel phenomena without Hermitian counterpart~\cite{Hatano1996,Hatano1998,Bener1998,Bener1999,Berry2004,Chong2010,Uzdin_2011,Heiss_2012,Liertzer2012,Leykam2017,Shen2018,Yao2018,Lee2019}. Exceptional point is one example~\cite{Bener1999,Berry2004,Chong2010,Uzdin_2011,Heiss_2012}, where both eigenvalues and eigenvectors coalesce different from diabolical point in Hermitian system, where only eigenvalues coalesce and the eigenvectors can still be orthogonalized. It has been realized in photonic~\cite{Makris2008,Ruter2010,Peng2014,Regensburger2012,Hossein2014,Feng2014}, acoustic~\cite{Zhu2014,Fleury2015,Ding2016,Zhu2018}, quantum systems~\cite{Peng2016,Zhang2016,Xu2017,Ding2021,Xiao2021,Liang2023}, and demonstrated tremendous potential in the generation, control and transmission of waves.

Recently, Floquet driving has been used to engineer exceptional points~\cite{Moiseyev2011,Luo2013,Joglekar2014,Lee2015,Chitsazi2017,Xie2018,Wang2018,Li2019,Xiao2019,Zhang2020,Harter2021,Ding2021,Wang2023,Liu2024}, demonstrated its control on the spectrum, including the number of exceptional points and the position in parameter space. Both theory~\cite{Joglekar2014,Lee2015} and experiment~\cite{Chitsazi2017,Li2019,Wang2023,Liu2024} have shown exceptional points can be arrived by choosing a proper driving even with a vanishingly small gain and loss. This property makes it possible achieve exceptional points in low loss system, $e.g.$ photonic, mechanical and atomic systems. Besides, Floquet system is unusual for its novel dynamical structure in time evolution operator~\cite{Nathan_2015}, which makes it support anomalous Floquet topological states~\cite{Jiang2011,Rudner2013,Zhu2021,zhu2024} that not exist in static system. However, as we know there is no work discuss the dynamical structures of Floquet exceptional points.

In this work, we propose a new kind of exceptional points, called Floquet $\pi$ exceptional points, and show its eigenvector rotates on Bloch sphere and accumulate $\pi$ geometric phase in one time period. We also show such kind of exceptional points exist in a Floquet non-Hermitian bipartite lattice and their merging play an important role in the phase transition. Specifically, pairs of order-1/2 Floquet $\pi$ exceptional points exist in quasimomentum gap phases~\cite{Zurita2009,Mark2022,ren2024}. When we decrease the strength of gain and loss, the merging of exceptional points is allowed for the same dynamical structure. Conversely, when we increase the strength of gain and loss, the merging is forbidden for opposite dynamical structure. So there is a phase transition from quasimomentum gap phase to quasienergy gap phase~\cite{Yu2024} when decreasing gain and loss and no phase transition when increasing gain and loss. At the phase transition point, the two order-1/2 Floquet $\pi$ exceptional points merging to one order-1 Floquet $\pi$ exceptional point. We then show around the order-1 Floquet $\pi$ exceptional point, the system is perfect transparency.

\begin{figure}
	\includegraphics[width=\linewidth]{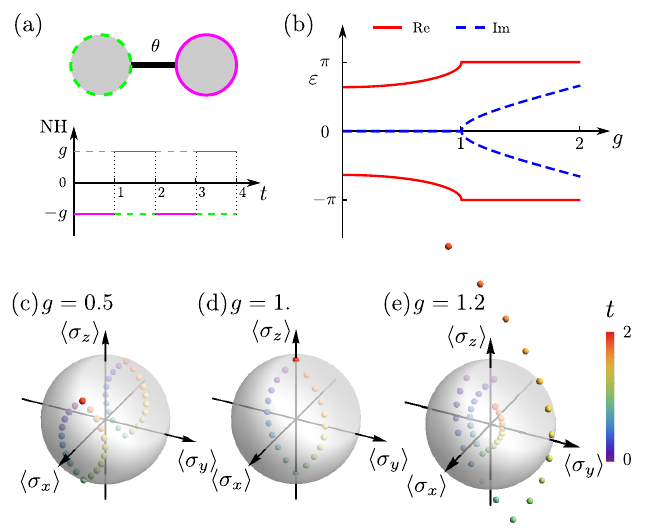}
	\caption{Floquet $\pi$ exceptional points in bi-resonators system with spatiotemporal modulated gain and loss. (a) Bi-resonators system. Two resonators are coupled by $\theta$. The modulation function for onsite non-Hermiticity of the two resonators are shown below. (b) Real part and imaginary part of quasienergy as function of $g$ with $\theta=1$. Trajectories of eigenvectors on Bloch sphere for $g=0.5$ (c), $g=1.$ (d) and $g=1.2$ (e), respectively.}
	\label{floquetpiEP}
\end{figure}

{\it Floquet $\pi$ Exceptional Point in bi-Resonators System.--} We first study the floquet $\pi$ exceptional points in a bi-resonators system. The model is shown in Fig.~\ref{floquetpiEP}(a), which contains two resonators with the onsite non-Hermiticity in each resonator is modulated in time as function shown in Fig.~\ref{floquetpiEP}(a). The time dependent Hamiltonian can be described by,
\begin{eqnarray}
	H(t)=\left\{
	\begin{array}{cc}
		H_1&\ell T<t\leq \ell T+T/2\;\\
		\sigma_x H_1\sigma_x& \ell T+T/2<t\leq \ell T+T\;
	\end{array}
	\right.
	\label{eq1}
\end{eqnarray}
where $\ell\in\mathbb{Z}$ and $H_1=\theta\sigma_x+ig\sigma_z$. Here, $\theta$ is the coupling between two resonators; $i$ is the imaginary unit; $g$ is the strength of gain; $\sigma_{x,y,z}$ are the Pauli matrices. $T$ is the driving period and we set $T=2$ in dimensionless units. The system is $\mathcal{PT}$ invariant at all instantaneous time, obeying $\mathcal{PT}H(t)\mathcal{PT}^{-1}=H(t)$, with $\mathcal{P}=\sigma_x$ the spatial inversion operator and $\mathcal{T}=K$ the spinless time reversal operator ($K$ is complex conjugate operator). Same Hamiltonian has been studied in integrated photonics recently~\cite{Liu2024}.

The eigenmodes $|\psi\rangle$ of periodically driven system can be obtained by solving eigenequation $U_T|\psi\rangle=e^{i\varepsilon}|\psi\rangle$, where $\varepsilon$ is quasienergy and $U_T\equiv \mathbb{T}\mathrm{exp}\big[\!-\!i\!\int_{0}^{T} H(\tau)\,d\tau\big]$ is the Floquet operator. $\mathbb{T}$ is the time-ordering operator. In the bi-resonators system, the Floquet operator can be simplified as 

\begin{equation}
	U_T=\sigma_xS\sigma_xS
\end{equation}
with $S=e^{iH_1}$. In Fig.~\ref{floquetpiEP}(b), we show the quasienergies as a function of $g$. When $g$ is smaller than 1, the quasienergies are real and the system is in $\mathcal{PT}$-unbroken phase. When $g$ is larger than 1, the quasienergies are complex and the system is in $\mathcal{PT}$-broken phase. At phase transition point $g=1$, the quasienergies coalesce to $2\ell\pi+\pi$ and the eigenvectors coalesce to $(1,0)$, which is main feature of exceptional points where both eigenenergies and eigenvectors coalesce. Although nonzero energy exceptional points can also be realized in static $\mathcal{PT}$ system by a whole energy shift, we will show here the Floquet $\pi$ exceptional points are nontrivial by novel dynamical properties.

The evolution of eigenvectors in one time period can be calculated by $|\psi(t)\rangle=U_t|\psi\rangle$, where $U_t \equiv \mathbb{T}\mathrm{exp}\big[\!-\!i\!\int_{0}^{t} H(\tau)\,d\tau\big]$ is time evolution operator. $|\psi(t)\rangle$ is also eigenvector of the system but for different time frame. We can show the dynamical evolution of eigenvectors on Bloch sphere by defining $\langle\sigma_x\rangle(t)$, $\langle\sigma_y\rangle(t)$ and $\langle\sigma_z\rangle(t)$ as 
\begin{subequations}
	\begin{eqnarray}
		\langle\sigma_x\rangle(t) =\langle\psi(t)|\sigma_x|\psi(t)\rangle\\
		\langle\sigma_y\rangle(t) =\langle\psi(t)|\sigma_y|\psi(t)\rangle\\
		\langle\sigma_z\rangle(t) =\langle\psi(t)|\sigma_z|\psi(t)\rangle
	\end{eqnarray}
\end{subequations}

Figs.~\ref{floquetpiEP}(c)-\ref{floquetpiEP}(e) show trajectories of eigenvectors for $\mathcal{PT}-$unbroken phase [Fig.~\ref{floquetpiEP}(c)], Floquet $\pi$ exceptional points [Fig.~\ref{floquetpiEP}(d)] and  $\mathcal{PT}-$broken phase [Fig.~\ref{floquetpiEP}(e)]. In the $\mathcal{PT}-$unbroken phase, there are two separate eigenvectors. Although not orthogonal due to non-Hermiticity, they evolve on different hemispheres, one on $x+$ hemisphere and one on $x-$ hemisphere. After one time period, the eigenvectors come back to itself. In $\mathcal{PT}-$broken phase, there are also two separate eigenvectors. They evolve on $yz-$plane. One amplitude increase and the other decrease as time evolve. At Floquet $\pi$ exceptional points, two eigenvectors coalesce to one. The eigenvector evolves on $yz-$plane. In one time period, the eigenvector comes back to itself and accumulates $\pi$ geometric phase. It is the reason we call it Floquet $\pi$ exceptional points but not the $\pi$ quasienergies which can be easily changed by quasienergy shift without change the dynamical properties.

\begin{figure}
	\includegraphics[width=\linewidth]{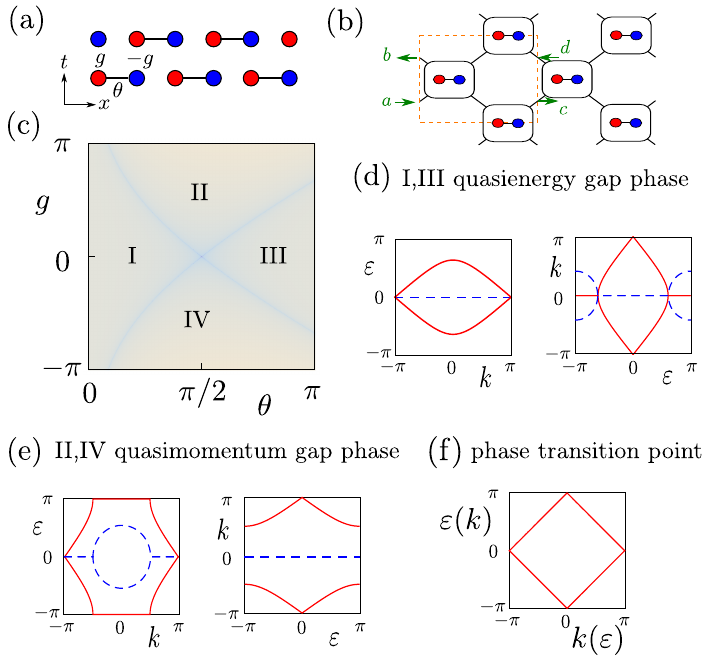}
	\caption{Floquet bipartite lattice model and phase diagram. (a) Floquet bipartite lattice under a two-step driving protocol. Red (blue) sites represent gain (loss) sites. Black lines represent nearest-neighbor couplings. (b) The equivalent nonunitary circuit model. Orange dashed line box marks the unit cell. (c) Phase diagram as function of $\theta$ and $g$. The parameter space is divided into four regions, marked by I, II, III and IV. I and III (II and IV) belong to quasienergy band gap phases (quasimomentum band gap phases). (b) Example of dual bands for I/III ($\theta=0.3\pi$, $g=0.2\pi$.) (c) Example of dual bands for II/IV ($\theta=0.3\pi$, $g=0.5\pi$). (d) Example of dual bands at phase transition line ($\theta=0.3\pi$, $g=1.1171$). Red solid (blue dashed) line represents real part (imaginary part) for quasienergy or quasimomentum.}
	\label{phasediagram}
\end{figure}

{\it Floquet Bipartite Lattice Model.--} We then study Floquet $\pi$ exceptional point in a bipartite lattice model as shown in Fig.~\ref{phasediagram}(a), which is periodic both in time and space. The unit cell of the system contains two sub-lattices, one is gain sub-lattice (red) and the other is loss sub-lattice (blue). Gain and loss are alternatively distributed both in space and time. Along time direction, the system contains two steps. The time dependent Bloch Hamiltonian in one time period can be described by piecewise function,
\begin{eqnarray}
	H(k,t)=\left\{
	\begin{array}{cc}
	H_1&\ell T<t\leq \ell T+T/2\;\\
	\sigma_{k/2} H_1\sigma_{k/2}& \ell T+T/2<t\leq \ell T+T\;
    \end{array}
	\right.
	\label{eq5}
\end{eqnarray}
where $\sigma_{k/2}$ is defined as $\sigma_{k/2}=e^{ik/2}\sigma_++e^{-ik/2}\sigma_-$ with $\sigma_\pm=(\sigma_x\pm i\sigma_y)/2$. Without gain and loss ($g=0$), the model is Floquet Su-Schrieffer-Heeger model, which has been shown to support fruitful topological phases~\cite{Dal2015,Asb2013,Asb2014,Raditya2018}, and can be realized in different systems~\cite{Zhou2017,Flurin2017,Cardano2017,Cheng2019,Xie2020}. 

The system is periodic along space and time direction, both quasienergy and quasimomentum are good quantum number. Its quasienergy band $\varepsilon(k)$ and eigenstates $|\psi_k\rangle$ can be obtained by solving eigen equation $U_T(k)|\psi_k\rangle=e^{i\varepsilon(k)}|\psi_k\rangle$, where $U_T(k)$ is Floquet operator expressed as 
\begin{equation}
	U_T(k)=\sigma_{k/2}S\sigma_{k/2}S
\end{equation}

The system can be equivalent to a nonunitary circuit model~\cite{zhu2024} as shown in Fig.~\ref{phasediagram}(b). Input and output signals at both ends of one unit cell, $(c,d)^{\text{T}}$ and $(a,b)^{\text{T}}$ are related by transfer matrix,

\begin{equation}
	M(\varepsilon)=\sigma_{\varepsilon/2}\tilde{S}\sigma_{\varepsilon/2}\tilde{S}
\end{equation}
where  $\sigma_{\varepsilon/2}$ is defined as $\sigma_{\varepsilon/2}=e^{i\varepsilon/2}\sigma_++e^{-i\varepsilon/2}\sigma_-$. And $\tilde{S}$ is a two-by-two matrix obtained from $S$ by $\tilde{S}_{11}=-S_{11}/S_{12}$, $\tilde{S}_{12}=1/S_{12}$, $\tilde{S}_{21}=-1/S_{12}$ and $\tilde{S}_{22}=-S_{122}/S_{12}$. A quasimomentum band $k(\varepsilon)$ can be obtained by solving equation $\det[M(\varepsilon)-e^{ik}]=0$. We call $\varepsilon(k)$ and $k(\varepsilon)$ dual bands.

{\it Phase Diagram.--}The phase diagram of the system in parameter space $(\theta, g)$ is shown in Fig.~\ref{phasediagram}(c), with $\theta\in[0,\pi]$ and $g\in[-\pi,\pi]$. We notice the parameter space is divided into four regions (I, II, III and IV) by two phase transition lines determined by

\begin{equation}
	\frac{g^2-\theta^2\cos(2\sqrt{\theta^2-g^2})}{g^2-\theta^2}=-1
\end{equation}

I and III belong to quasienergy band gap phases, while II and IV belong to quasimomentum band gap phases. Fig.~\ref{phasediagram}(d) show one example of dual bands for quasienergy gap phases, where we notice the quasienergy band is in $\mathcal{PT}$-unbroken phase and the quasimomentum band is in $\mathcal{PT}$-broken phases. Fig.~\ref{phasediagram}(e) show one example for quasimomentum gap phases, where we notice the quasienergy band is in $\mathcal{PT}$-broken phase and the quasimomentum band is in $\mathcal{PT}$-unbroken phases. At phase transition point, the dual bands are the same, with gapless and linear dispersion as shown in Fig.~\ref{phasediagram}(f).  

{\it Dynamical Properties and Merging of Floquet $\pi$ Exceptional Points.--}The quasienergy band for quasimomentum gap phase has two Floquet $\pi$ exceptional points at $k=\pm k_c$ with $k_c=\arccos[-1+(-2+2g^2/\theta^2)\text{csch}^2(\sqrt{g^2-\theta^2})]$. These exceptional points at $\varepsilon=\pi$ have unusual dynamical properties. 

\begin{figure}
	\includegraphics[width=\linewidth]{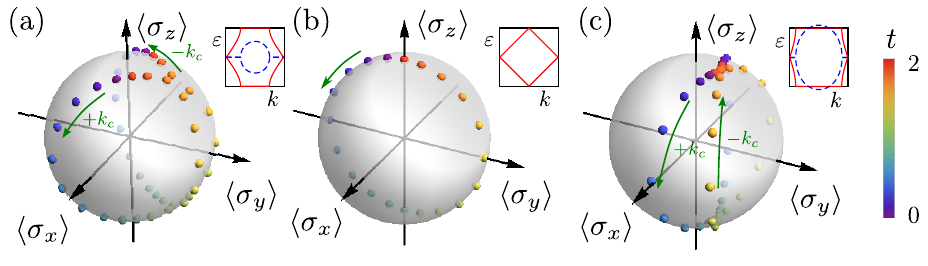}
	\caption{The trajectory of the eigenvector of $\pi$-type exceptional point on Bloch sphere. (a) for quasimomentum gap phase ($\theta=0.3\pi$ and $g=0.5\pi$). (b) for a phase transition point ($\theta=0.3\pi$ and $g=1.1171$). (c) also for quasimomentum gap phase but with larger gain and loss ($\theta=0.3\pi$ and $g=\pi$). Here the eigenvectors are normalized at each instantaneous time in plotting.}
	\label{piEP}
\end{figure}

Fig.~\ref{piEP}(a) show one example for $\theta=0.3\pi$ and $g=0.5\pi$ where system belongs to quasimomentum gap phases. We notice two Floquet $\pi$ exceptional points rotate along different axis. They rotate once and accumulate $\pi$ geometric phase in one time period. When we slightly decrease gain and loss, the momentum of Flqouet $\pi$ exceptional points slightly moves toward $k=0$ and their trajectories on Bloch sphere prefer to rotate along $x$ axis in anticlockwise direction. At critical value $g\approx1.1171$, two Floquet $\pi$ exceptional points merge to one as shown in Fig.~\ref{piEP}(b). Conversely, when we increase gain and loss, the two Floquet $\pi$ exceptional points trend to $k=\pm\pi$ points and rotate along $y-$axis as shown in Fig.~\ref{piEP}(c). However, the two of them rotates clockwise and counterclockwise, respectively. Therefore, they can only approach $\pm\pi$ infinitely, but will not merge into one.

\begin{figure}
	\includegraphics[width=\linewidth]{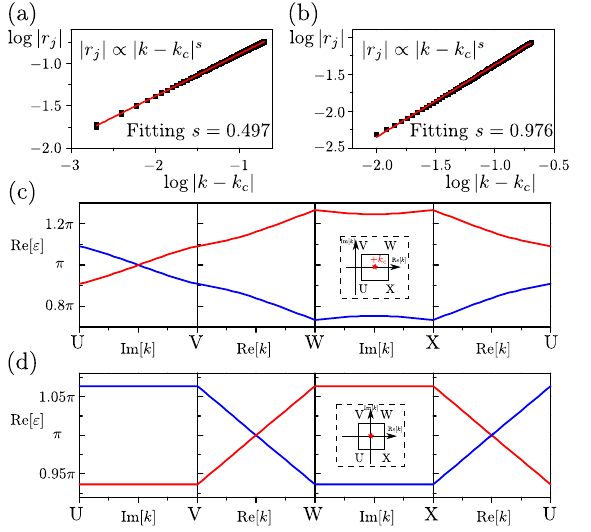}
	\caption{Phase rigidity around Floquet $\pi$ exceptional points. (a) Log-log plot of phase rigidity $|r_j|$ versus $|k-k_c|$ for Floquet $\pi$ exceptional point in Fig.~\ref{piEP}(a). (b) Same as (a) but for Floquet $\pi$ exceptional point in Fig.~\ref{piEP}(b). (c) Quasienergy trajectories for looping around Floquet $\pi$ exceptional point in (a). (d) Same as (c) but for Floquet $\pi$ exceptional point in (b).}
	\label{orderEP}
\end{figure}

To show more details of Floquet $\pi$ exceptional points merging, we study phase rigidity~\cite{Rotter_2009,Ding2016} of system which is defined as
\begin{equation}
	r_j=\frac{\langle\tilde{\psi}^{j}_{k}|\psi^{j}_{k}\rangle}{\langle\psi^{j}_{k}|\psi^{j}_{k}\rangle}
\end{equation}
$|\psi^{j}_{k}\rangle$ and $|\tilde{\psi}^{j}_{k}\rangle$ are the $j^{th}$ right and left eigenstates of $U_T(k)$, obtained from $U_T(k)|\psi^{j}_{k}\rangle=e^{i\varepsilon_{j}}|\psi^{j}_{k}\rangle$ and $\langle\tilde{\psi}^{j}_{k}|U_T(k)=e^{i\varepsilon_{j}}\langle\tilde{\psi}^{j}_{k}|$, respectively. Fig.~\ref{orderEP}(a) show log-log plot of phase rigidity $|r_j|$ versus $|k-k_c|$ around Floquet $\pi$ exceptional points for quasimomentum gap phase, from which we find the exponent is 1/2, a signature of order-1/2 exceptional point but here is Floquet $\pi$ one. The order-1/2 exceptional point can also be confirmed by performing adiabatic process circling in the complex $k$ plane as shown in Fig.~\ref{orderEP}(c). We notice when $k$ rotates around exceptional point one circle, two states exchange positions with each other. The behavior of Floquet $\pi$ exceptional point at phase transition point from merging of two order-1/2 exceptional points is totally different. Fig.~\ref{orderEP}(b) show the log-log profile at phase transition point, where we find the exponent is 1, a signature of order-1 exceptional points from merging two order-1/2 exceptional points with same charges~\cite{Shen2018,Ding2016}. Different from the order-1/2 exceptional points, here when $k$ rotates around order-1 exceptional point one circle, the states come back to itself as shown in Fig.~\ref{orderEP}(d). So the phase transition line in Fig.~\ref{phasediagram}(c) is actually composed of a class of order-1 Floquet $\pi$ exceptional points.

{\it Scattering Properties of Order-1 Floquet $\pi$ exceptional point.--}
The previous text has proven that the system supports order-1 Floquet $\pi$ exceptional point at phase transition point, next we study its scattering properties. For the Floquet bipartite lattice, the transfer matrix for $N$ unit cells is expressed as

\begin{equation}
	\mathcal{M}=[\sigma_{\varepsilon/2}\tilde{S}]^{2N}
\end{equation}

The spectral transmission and reflection coefficients for left ($l$) and right ($r$) incidence can be expressed in terms of the transfer matrix elements as
\begin{equation}
t^{(l)}=t^{(r)}\equiv t=\frac{1}{\mathcal{M}_{22}}, r^{(l)}=-\frac{\mathcal{M}_{21}}{\mathcal{M}_{22}}, r^{(r)}=\frac{\mathcal{M}_{12}}{\mathcal{M}_{22}}
\end{equation}

\begin{figure}
	\includegraphics[width=\linewidth]{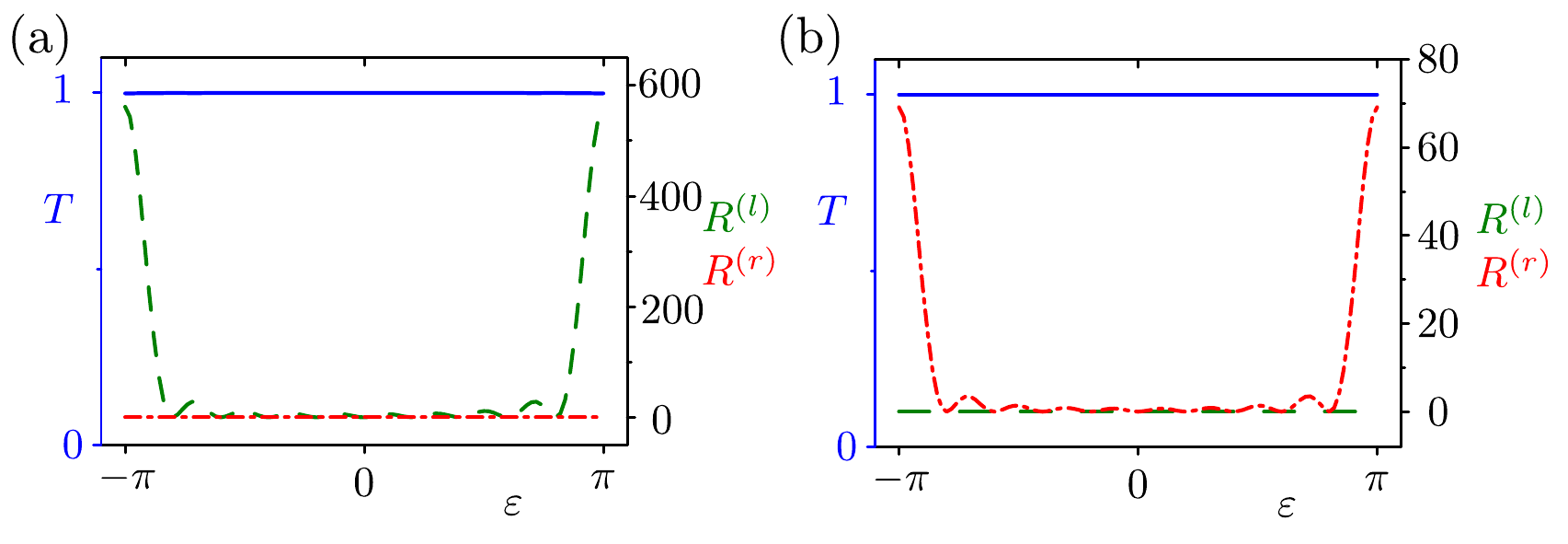}
	\caption{Spectral density transmission and reflection at phase transition point. (a) for a phase transition point wit $\theta=0.3\pi$ and $g=1.1171$. (b) for a phase transition point wit $\theta=0.7\pi$ and $g=0.9148$. }
	\label{Transmission}
\end{figure}

Fig.~\ref{Transmission} show the spectral density transmission and reflection (defined as $T=|t|^2$, $R^{(l)}=|r^{(l)}|^2$ and $R^{(r)}=|r^{(r)}|^2$) around two Floquet $\pi$ exceptional points. We notice the system is totally transparency in both cases, with unit transmission for full quasienergies. For the other Floquet $\pi$ exceptional points on the phase transition line in Fig.~\ref{phasediagram}(c), the system is also transparency and we will not show it again. The non-Hermitian properties of the system can be seen from the reflection of the system. We notice the reflections for left incidence and right incidence are different, with largest difference at the order-1 Floquet $\pi$ exceptional point. For $\theta=0.3\pi$ and $g=1.1171$, the left-incidence reflection is large and the right-incidence reflection is zero as shown in Fig.~\ref{Transmission}(a). While for another case $\theta=0.7\pi$ and $g=0.9148$, things are different where the right-incidence reflection is large and the left-incidence reflection is zero as shown in Fig.~\ref{Transmission}(b). Such difference can be understand by parameter evolution in $(\theta, g)$, $e.g.$ $(0.3\pi,1.1171)\rightarrow(0.3\pi,0)\rightarrow(0.7\pi,0)\rightarrow(0.7\pi,0.9148)$. In the middle process from $(0.3\pi,0)$ to $(0.7\pi,0)$, band inversion happens at $(0.5\pi,0)$, which changes the physical properties of Floquet $\pi$ exceptional point. Need to say, such kind of difference can also be realized by exchanging gain and loss, $e.g.$ $(0.3\pi,1.1171)\rightarrow(0.3\pi,-1.1171)$.

{\it Conclusion.--} Exceptional points in Floquet system can have totally different behavior as comparing with those in static system. The difference comes from the dynamical structure in the time evolution operators. The Floquet $\pi$ exceptional points we find here is characterized by rotating eigenvectors on Bloch sphere, and plays an important role for the phase transition between quasimomentum gap phase and quasienergy gap phase by merging of two order-1/2 Floquet $\pi$ exceptional points to one order-1 Floquet $\pi$ exceptional point. The merging of Floquet $\pi$ exceptional points is constrained by their dynamical structure. At the phase transition point between quasimomentum gap phase and quasienergy gap phase, the quasienergy band structure of the system is real and linear in full bands, and degenerate at ($\varepsilon=\pi, k=0$) like a diabolical point in Hermitian system. However, we show their eigenvectors are different. Those degenerate points are order-1 Floquet $\pi$ exceptional points. Their spectral transmission is unit and the reflection from one side is obvious larger than the other side.

\section{Acknowledgement}
W. Z. acknowledges support from the Start up Funding from Ocean University of China.

\end{document}